\documentclass[conference]{IEEEtran}
\IEEEoverridecommandlockouts
\usepackage{cite}
\usepackage{amsmath,amssymb,amsfonts}
\usepackage{algorithmic}
\usepackage[caption=false]{subfig}
\usepackage[hidelinks, bookmarks=false]{hyperref}
\hypersetup{
    colorlinks=true,
    linkcolor=black,     
    urlcolor=blue,
    citecolor=black
    }
\usepackage{graphicx}
\usepackage{fancyhdr}
\usepackage{pgfplots}
\pgfplotsset{compat=1.18} 
\usepackage{textcomp}
\usepackage{float}
\usepackage{xcolor}
\def\BibTeX{{\rm B\kern-.05em{\sc i\kern-.025em b}\kern-.08em
    T\kern-.1667em\lower.7ex\hbox{E}\kern-.125emX}}

\fancypagestyle{IEEEpubidstyle}{
  \fancyhf{}

  \fancyfoot[L]{979-8-3315-2103-5/24/\$31.00~\copyright2024 IEEE\\ \textcolor{red}{\footnotesize S. Viththarachchige, D. Alexander, S. Rajendran and V. Aravinthan, "Social Equity Based Optimal Power Flow Framework to Hedge Against Price Events," 2024 56th North American Power Symposium (NAPS), El Paso, TX, USA, 2024, pp. 1-6, \textbf{doi: 10.1109/NAPS61145.2024.10741650}.}}
}

\begin{document}

\title{Social Equity Based Optimal Power Flow Framework to Hedge Against Price Events
\thanks{This work was sponsored by US NSF EPSCoR RII Track 1 \#2148878}
}
\author{\IEEEauthorblockN{Sachinth Viththarachchige, Demy Alexander, Sarangan Rajendran and Visvakumar Aravinthan}
\IEEEauthorblockA{\textit{Department of Electrical and Computer Engineering} \\
\textit{Wichita State University}\\
Wichita, KS, USA\\
\href{mailto:syviththarachchige@shockers.wichita.edu}{syviththarachchige@shockers.wichita.edu}}

}

\maketitle

\thispagestyle{IEEEpubidstyle}

\begin{abstract}
With the increasing frequency of high impact low probability events, electricity markets are experiencing significant price spikes more often. This paper proposes a novel social equity driven optimal power flow framework to mitigate the adverse effects of price events that lead to such price spikes. The framework integrates social welfare optimization with socioeconomic considerations by including a socioeconomic score that quantifies the energy burden and socioeconomic status of consumers. By incorporating both supply cost and consumer satisfaction, the model aims to achieve a balanced and fair distribution of resources during price events, while considering resource scarcity and possible load curtailment. The proposed framework is tested for convergence on modified versions of the PJM 5-bus system and IEEE 24-bus reliability test system, discussing its potential effectiveness in enhancing social equity and optimizing power flow under system security constraints. Sensitivity analysis further highlights the impact of socioeconomic score on social welfare, providing insights for future improvements.
\end{abstract}


\section{Introduction}
From natural and man-made high impact low probability (HILP) phenomena, the frequency of price events is becoming considerably high. Price events result in price volatility that leads to exceedingly high electricity prices in day-ahead and real-time electricity markets. Some of the most common price events are, natural disasters, renewable energy volatility, line congestion, and cyber attacks \cite{eventsintro}. 

\begin{figure}[b!]
\centerline{\includegraphics[width=\linewidth]{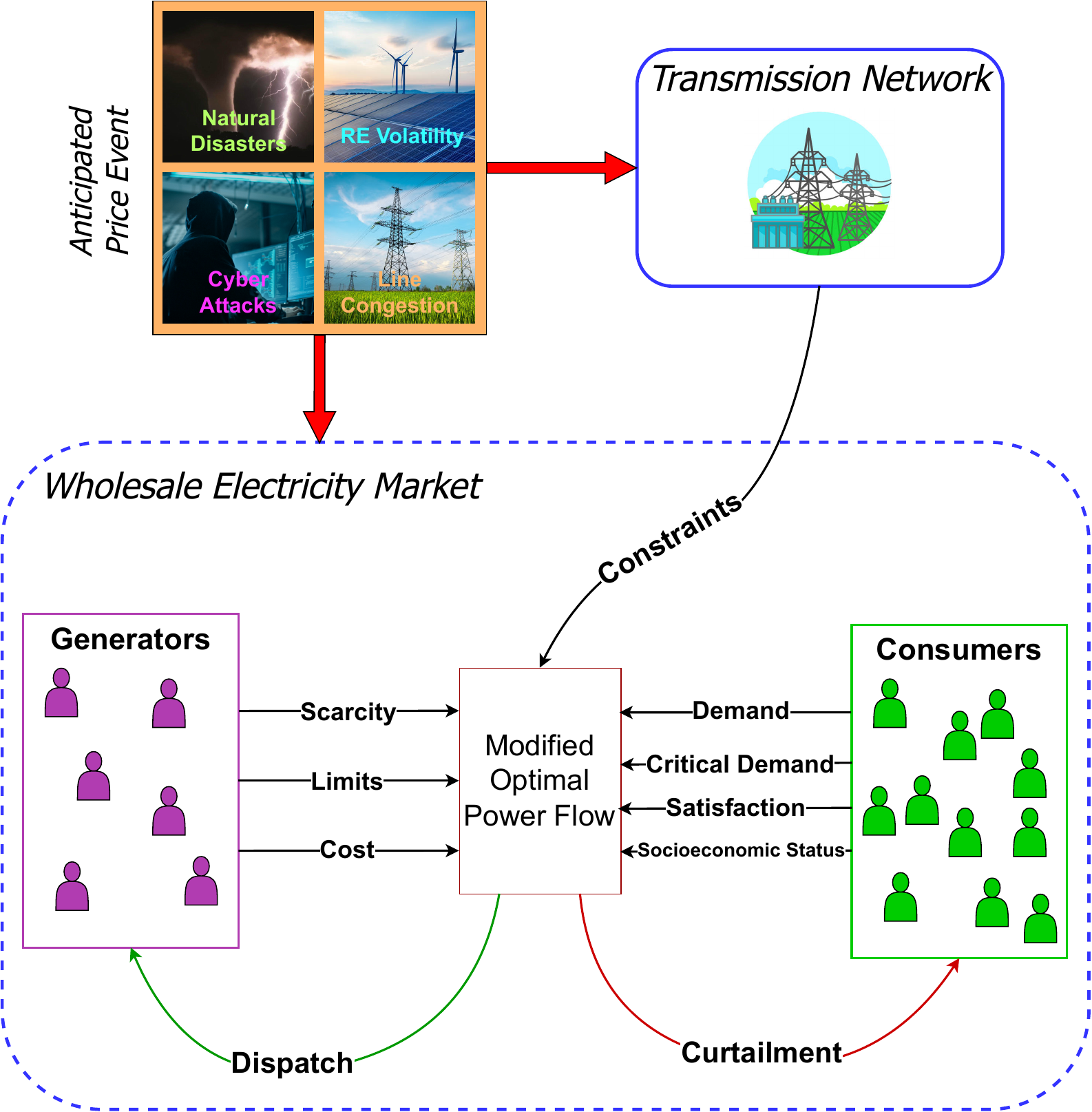}}
\caption{Proposed social equity based optimal power flow framework}
\label{priceevents}
\end{figure}

A major focus on price events has emerged from recent events such as, a polar vortex accompanying winter storms including North American winter storm Uri (2021), and winter storm Elliott (2022). During a winter storm caused by polar vortex, the energy reliability council of Texas (ERCOT) had artificially inflated the electricity prices up to as high as \$9000/MWh for several days \cite{vortex}. Another price event situation was created by Uri to an extent that resulted in locational marginal prices (LMP) above \$3000/MWh at some LMP nodes in the day-ahead market of the grid operator, Southwest power pool (SPP) \cite{spp}. In a similar manner, winter storm Elliott led to price events among several regional transmission organizations (RTO) and independent system operators (ISO) \cite{ferc}.

Key observations of the price-spikes during extreme price events include, but not limited to \cite{spp}:
\begin{itemize}
    \item \textit{Unavailable generation and fuel}: Lack of fuel often leads to scarcity in generation.
    \item \textit{High gas prices}: Extremely high natural gas prices often are the primary drivers of record-high offers in the wholesale electricity markets.
    \item \textit{Increased credit exposure}: Spikes in market prices create liquidity issues to market participants and put lenders and investors at risk.
    \item \textit{Congested transmission}: Line congestion significantly increases in cases of loss of transmission assets.
\end{itemize}

Price events raise concerns that revolve around both supply side participants and demand side participants in a wholesale electricity market. It is important to take a holistic approach to tackle the adversaries that rise from price events. To achieve a fairness for both parties in a wholesale market experiencing price events, the combination of the technical aspect from power systems (cost-centric) and the social welfare aspect from welfare economics \cite{welfareecon} is required. 

\subsection{Related Work}
The inclusion of social welfare concepts in power system optimization has been there since early 1980s \cite{swofirst}. This was primarily done by addition of the concept of consumer utility or benefit \cite{utilfunc} to the standard system cost optimization problem. Initial work were more centered around long term planning problems as in \cite{swofirst}. With the introduction and adoption of demand response (DR) programs in electricity markets, subsequent studies have considered including price dependent loads (i.e., dispatchable loads) in the optimal power flow (OPF) problem \cite{weber}.

Reference \cite{swoeqn} proposes a fully decentralized social welfare optimization formulation that uses a multi-agent approach. They claim that by modeling the grid in a way that allows both generators and loads to be adjustable, economic dispatch problem and the DR problem can be integrated effectively. 

Reference \cite{p2p} introduces social welfare optimization to peer-to-peer (P2P) energy markets built upon community grids. It extends the price dependent consumer model concept to P2P energy trading that involves prosumers.

However, none of the existing literature have modeled an OPF formulation that specifically addresses resource scarcity (generator scarcity and/or line congestion) scenarios along with load curtailment, via social welfare optimization. Such a formulation could naturally become an ideal candidate for an OPF problem that attempts to tackle situations caused by price events.

Even though there are numerous standard market instruments to handle price events such as, virtual bidding, reserve markets, and flexible ramping products \cite{eventsintro},~\cite{econbook}, they do not usually consider the average energy burden or other socioeconomic factors of market participants.

\subsection{Proposed Framework}
We propose a social equity driven OPF framework to be used during price events in wholesale electricity markets. The key contribution of this work is formulation of a unified social welfare based optimization model that considers both supply cost and satisfaction of consumers (willingness to pay) in the face of resource scarcity during price events. Inclusion of social equity factor is achieved via a metric that quantifies the average energy burden and other factors of socioeconomic status of consumers. The aim is to maximize social welfare in case of potentially extreme LMPs experienced by the market participants, considering their socioeconomic status and price sensitivities.

The paper is organized as follows. Section II introduces the formulation of the model with required concepts from welfare economics and power systems, giving modeling assumptions. Section III explains the implementation of the model in test systems and the analysis approach. Section IV presents the results with their interpretation. Finally, Section V conveys concluding remarks based on implication of results.
\section{Model Development}
\subsection{Consumer Model of Aggregators}
From the perspective of an ISO or RTO, a load aggregator (from here onwards, will be referred to as aggregator) can be seen as a consumer in the wholesale electricity market. It is assumed that the ISO/RTO has exogenous information about every aggregator. These include purchase behavior, price sensitivity, and load profile during normal operation and during price events.

Due to the constraints imposed by the physical system, there may be a need for load curtailment during price events. This may include both voluntary and involuntary load curtailment \cite{nirasha} of downstream consumers. With this possible adjustment of loads, the aggregator is assumed to be a dispatchable load.

\subsection{Consumer Satisfaction Function}
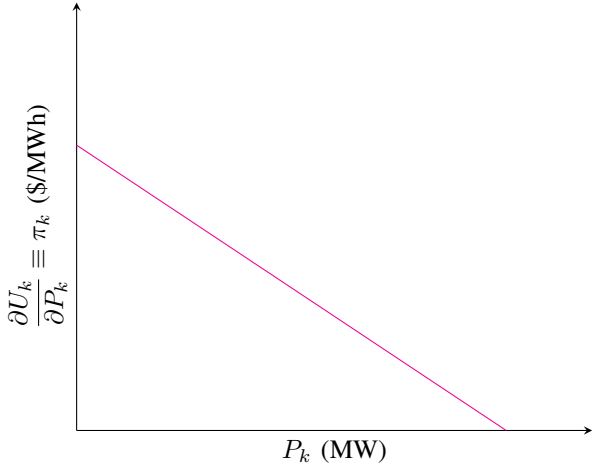
\begin{figure}[t!]
\centering
\begin{tikzpicture}
\begin{axis}[
    axis lines = left,
    xlabel = $P_k$~(MW),
    ylabel = $\dfrac{\partial U_{k}}{\partial P_{k}}\equiv\pi_k$~(\$/MWh),
    ticks=none,
    xmin=0, xmax=12,
    ymin=0, ymax=15
]
\addplot[
    domain=0:10, 
    samples=100, 
    color=magenta,
]
{-x + 10};
\end{axis}
\end{tikzpicture}
\caption{Inverse demand function of a transmission level consumer}
\label{inv_dem}
\end{figure}

A consumer's satisfaction or the willingness to pay for a commodity can be evaluated using the consumer benefit function from welfare economics theory\cite{utilfunc}. Several authors have used the same analogy in their efforts to develop frameworks that incorporate social welfare into the power systems operation and electricity markets \cite{satiscase1},~\cite{satiscase2}.
In this work, we shall call it the satisfaction function. The satisfaction of an aggregator $k$, consuming power $P_{k}$ can be given as: 

\begin{equation}
U_{k}(P_{k}) = 
\begin{cases}
\begin{aligned}
&\gamma_{k}P_{k}-0.5\mu_{k}P_{k}^2, \qquad\text{if~}~ 0\leq P_{k} \leq \dfrac{\gamma_{k}}{\mu_{k}}\\
&0.5\dfrac{\gamma_{k}^{2}}{\mu_{k}}, \qquad\text{if~}~ P_{k} \geq \dfrac{\gamma_{k}}{\mu_{k}}
\end{aligned}
\end{cases}
\end{equation}
where the parameters $\gamma_{k}$ and $\mu_{k}$ are defined by the load profile and the purchase behavior of the aggregator \cite{swoeqn}. The function is more likely to stay within the concave, non-linear region during price events, due to possible load curtailments, which leads to reduction of $P_{k}$. An interesting observation from this function is that $\frac{\partial U_{k} (\bullet)}{\partial P_k}$ gives the inverse demand function for the price $\pi_{k}$ which is depicted in Fig.~\ref{inv_dem}. The graph is monotonically decreasing due to the assumed concavity of $U_{k}$. By flipping the axes of the same graph (i.e., taking the inverse of the function $\pi_{k}$), the demand of an aggregator can be represented as a function of the electricity price \cite{weber}.

These relationships imply the price sensitivity of an aggregator participating in a wholesale electricity market. With this consideration, a demand bus of a transmission network with several aggregators can be modeled as price sensitive consumers, each having their own satisfaction function as shown in Fig.~\ref{concept_fig}. As mentioned earlier, the ISO/RTO is assumed to be well aware of all these information of each aggregator participating in the wholesale market.

\subsection{Socioeconomic Score}
\begin{figure}[t!]
\centerline{\includegraphics[width=\linewidth]{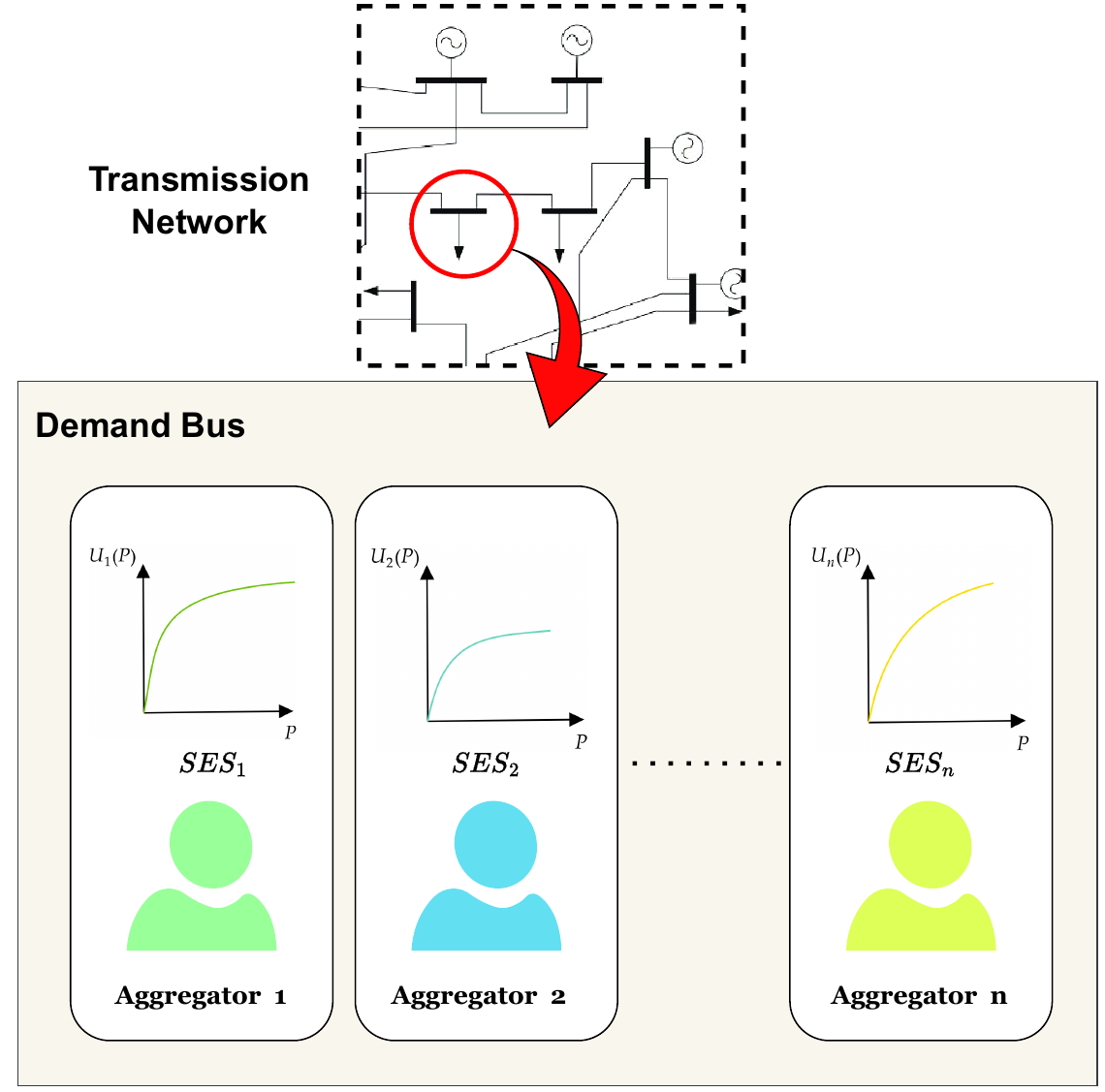}}
\caption{Representation of aggregators of a demand bus}
\label{concept_fig}
\end{figure}

To incorporate the social equity aspect, we add a metric called socioeconomic score (SES) to the model. This is a non-negative scalar value which represents the socioeconomic status \cite{sociofac} of communities that are being served by the aggregator. We use the convention such that, an aggregator serving a region with a high average energy burden and a low socioeconomic status would be assigned a high SES, and vice versa. Each aggregator has an SES as indicated in Fig.~\ref{concept_fig}.

\subsection{Objective Function}
The objective function of our model is twofold. It has a form similar to previous social welfare optimization work in power systems. The first part is the summation of satisfaction functions of aggregators, each weighted by their SES $\sigma_{d,a}$. And the remaining is the generator cost counterpart of the objective.
\begin{equation}
    \max\quad \sum_{d\in \mathbb{\mathcal{D}}}\sum_{a\in \mathbb{\mathcal{A}}_d} \sigma_{d,a} U_{d,a}(P_{d,a}) - \sum_{g\in \mathbb{\mathcal{G}}} C_{g}(P_g)
    \label{objFun}
\end{equation}

\begin{equation}
    \sum_{a\in \mathbb{\mathcal{A}}_d}P_{d,a} = P_d
\end{equation}

\begin{equation}
    C_g(P_{g}) = a_{g}P_{g}^2+b_{g}P_{g}+c_{g}
\end{equation}
where, $\mathbb{\mathcal{D}}$ is the set of demand buses indexed by $d$, $\mathbb{\mathcal{G}}$ is the set of generator buses indexed by $g$, and $\mathbb{\mathcal{A}}_{d}$ is the set of aggregators at the demand bus $d$, indexed by $a$. $U_{d,a}(\bullet)$, $P_{d,a}$, and $P_{g}$ are satisfaction function, active power demand of the $a$\textsuperscript{th} aggregator of the $d$\textsuperscript{th} demand bus, and the active power output of the $g$\textsuperscript{th} generator, respectively. $a_{g}$, $b_{g}$, and $c_{g}$ are the cost coefficients of the quadratic cost function of the $g$\textsuperscript{th} generator.

\subsection{Constraints}
\subsubsection{Power Flow Equations}
Active and reactive power flow equations (\ref{peqn}) and (\ref{qeqn}) apply to all buses except the slack bus denoted by $i_{\text{slack}}$.
\begin{align}
    &\sum_{g\in\mathbb{\mathcal{G}}(i)}P_{g}-\sum_{d\in \mathbb{\mathcal{D}}(i)}P_{d}=V_{i}\sum_{j\in \mathbb{\mathcal{B}}}V_{j}(G_{ij}\cos\theta_{ij}+B_{ij}\sin\theta_{ij}),\notag\\ &\qquad\forall i\in\mathbb{\mathcal{B}}\setminus\{ i_{\text{slack}}\}
    \label{peqn}
\end{align}
where $\mathbb{\mathcal{B}}$,~$\mathbb{\mathcal{G}}(i)$,~and $\mathbb{\mathcal{D}}(i)$ are the set of buses, the set of generators, and the set of total aggregator demands at the $i$\textsuperscript{th} bus, respectively. $V_{i}$ denotes the voltage magnitude at the $i$\textsuperscript{th} bus. $G_{ij}$ and $B_{ij}$ denote the real and imaginary parts of the element at the $i$\textsuperscript{th} row and $j$\textsuperscript{th} column of the bus admittance matrix, respectively. $\theta_{ij}$ is the difference between the angles of the $i$\textsuperscript{th} bus and $j$\textsuperscript{th} bus ($\theta_{i}-\theta_{j}$).
\begin{align}
    &\sum_{g\in\mathbb{\mathcal{G}}(i)}Q_{g}-\sum_{d\in \mathbb{\mathcal{D}}(i)}Q_{d}=V_{i}\sum_{j\in \mathbb{\mathcal{B}}}V_{j}(G_{ij}\sin\theta_{ij}-B_{ij}\cos\theta_{ij}),\notag\\ &\qquad\forall i\in\mathbb{\mathcal{B}}\setminus\{ i_{\text{slack}}\}
    \label{qeqn}
\end{align}
where $Q_{g}$,~and $Q_{d}$ are the reactive power output of the generator $g$, and the total reactive power demand of the aggregators at the demand bus $d$, respectively.

\subsubsection{Line Flow Limits}
Taking the real and imaginary parts of the admittance of a transmission line between buses $i$ and $j$ as $g_{ij}$ and $b_{ij}$ respectively, the power $P_{l}$ flowing through the line can be calculated as:
\begin{align}
    &P_l = P_{ij} = V_{i}^{2}g_{ij}-V_{i}V_{j}(g_{ij}\cos\theta_{ij}+b_{ij}\sin\theta_{ij})~,\\&\qquad\forall i,j\in \mathbb{\mathcal{B}},~\forall (i,j)\in \mathbb{\mathcal{L}},~i\neq j \notag
\end{align}
neglecting the charging current of the line. $\mathbb{\mathcal{L}}$ is the set of transmission lines of the power system. Then, the line flow is limited by $S_{l}^{\textsl{max}}$ as:
\begin{equation}
    P_{l}\leq S_{l}^{\textsl{max}},\qquad \forall l \in \mathbb{\mathcal{L}}
\end{equation}

\subsubsection{Voltage Magnitude Limits}
The magnitude of the voltage at each bus $i$ is constricted within the safe operating limits $V_{i}^{\textsl{min}}$ and $V_{i}^{\textsl{max}}$ as,
\begin{equation}
   V_{i}^{\textsl{min}}\leq V_{i}\leq V_{i}^{\textsl{max}},\qquad \forall i \in \mathbb{\mathcal{B}}
\end{equation}

\subsubsection{Generator Dispatch Limits}
Each generator $g$ has their active and reactive power outputs and limited as shown in (\ref{pglim}) and (\ref{qglim}),
\begin{equation}
   P_{g}^{\textsl{min}}\leq P_{g}\leq P_{g}^{\textsl{max}},\qquad \forall g \in \mathbb{\mathcal{G}}
   \label{pglim}
\end{equation}

\begin{equation}
   Q_{g}^{\textsl{min}}\leq Q_{g}\leq Q_{g}^{\textsl{max}},\qquad \forall g \in \mathbb{\mathcal{G}}
   \label{qglim}
\end{equation} \cite{opfbook}

\subsubsection{Aggregator Power Limits}
We define power limits (\ref{palimits}) and (\ref{qalimits}) for each aggregator, during price events.
\begin{itemize}
    \item Critical power limits:~Denoted by $P_{d,a}^{\textsl{c}}$, and $Q_{d,a}^{\textsl{c}}$, these lower bounds define the minimum allowable demands of the aggregator ($d,a$) after possible load curtailment. These value may be taken as the total of the most critical loads\cite{adi}.
    \item Normal power limits:~The upper bounds $P_{d,a}^{\textsl{n}}$ and $Q_{d,a}^{\textsl{n}}$ define the normal power demands of the aggregator ($d,a$) in absence of price events.
\end{itemize}
Both types of these power limits are assumed to be intelligently known by the ISO/RTO through forecasting. These values are different from case to case. Load curtailment of each aggregator is given by, $P_{d,a}^{\textsl{n}}-P_{d,a}$.
\begin{equation}
   P_{d,a}^{\textsl{c}}\leq P_{d,a}\leq P_{d,a}^{\textsl{n}},\qquad \forall d \in \mathbb{\mathcal{D}},~\forall a \in \mathbb{\mathcal{A}}_{d}
   \label{palimits}
\end{equation}
\begin{equation}
   Q_{d,a}^{\textsl{c}}\leq Q_{d,a}\leq Q_{d,a}^{\textsl{n}},\qquad \forall d \in \mathbb{\mathcal{D}},~ \forall a \in \mathbb{\mathcal{A}}_{d}
   \label{qalimits}
\end{equation}
There can be situations where $P_{d,a}^{\textsl{n}}-P_{d,a}\geq P_{d,a}^{\textsl{n}}-P_{d,a}^{\textsl{c}}$ (not being able to meet the critical load of certain aggregators, inevitably). In such cases, the lower bound constraint of such aggregators should be relaxed and the problem needs to be solved again in successive stages to remove the infeasibility. However, it is beyond the scope of this formulation study.

\subsubsection{Adequacy Constraints}
The constraints (\ref{adep}) and (\ref{adeq}) ensure the sufficient availability of generator power output to cater to the current aggregator demand levels.
\begin{equation}
    \sum_{g\in \mathbb{\mathcal{G}}}P_{g} \geq \sum_{d\in \mathbb{\mathcal{D}}}\sum_{a\in \mathbb{\mathcal{A}}_d}P_{d,a}
    \label{adep}
\end{equation}
\begin{equation}
    \sum_{g\in \mathbb{\mathcal{G}}}Q_{g} \geq \sum_{d\in \mathbb{\mathcal{D}}}\sum_{a\in \mathbb{\mathcal{A}}_d}Q_{d,a}
    \label{adeq}
\end{equation}

\begin{figure}[t!]
\centerline{\includegraphics[width=\linewidth]{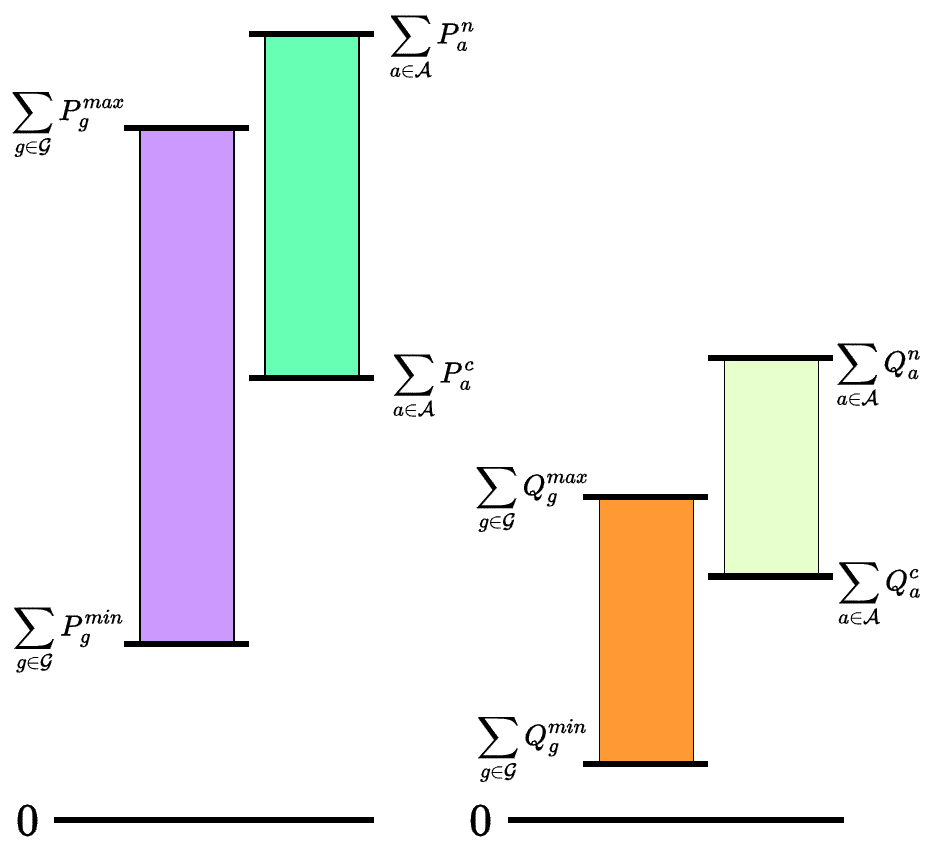}}
\caption{Illustration of generation scarcity during price events}
\label{resscar}
\end{figure}

The decision variables of the optimization problem are $P_{g}$,~$Q_{g}$,~$P_{d,a}$,~and $Q_{d,a}$, indexed by their respective sets. The other variables $\mathbf{V}$ and $\mathbf{\theta}$ can be considered as state variables. Fig.~\ref{resscar} illustrates the generation scarcity during a price event, where $\mathbb{\mathcal{A}}$ is the set of all aggregators in the system. Using a similar notation, the total effective load curtailment can be written as, $\sum_{g\in \mathbb{\mathcal{G}}}P_g - \sum_{a \in \mathbb{\mathcal{A}}}P_a$. Furthermore, it is possible to have line congestion concurrently.

\section{Case Study}
\subsection{Test Systems}
We implement the model using two selected test systems: a modified version of the 5-bus system from \cite{pjm5}, and a modified version of the IEEE 24-bus reliability test system \cite{ieee24}. To simulate a resource scarcity scenario, the normal power limits of the aggregators ($P_{d,a}^{\textsl{n}}$ and $Q_{d,a}^{\textsl{n}}$) are set such that, $\sum P_{d,a}^{\textsl{n}} \geq \sum P_{g}^{\textsl{max}}$ and $\sum Q_{d,a}^{\textsl{n}} \geq \sum Q_{g}^{\textsl{max}}$ in both systems. The ratings of the lines are reduced to $15$\% to $80$\% of their original values to add the congestion aspect. The parameters of the aggregators for the 5-bus system is shown in Table~\ref{aggrdata5}. New line ratings of the 5-bus system are shown in Table~\ref{linedata5}. In a similar manner, the parameters of the aggregators are assigned in the 24-bus system. The generator cost coefficients of the 5-bus system are modified as shown in Table~\ref{gendata5}. The generator cost coefficients of the 24-bus system are unchanged. Every demand bus in both systems has $2$ to $3$ aggregators with their attributes, $\sigma_{d,a}$,~$\gamma_{d,a}$,~$\mu_{d,a}$,~$P^{\textsl{n}}_{d,a}$,~$P^{\textsl{c}}_{d,a}$,~$Q^{\textsl{n}}_{d,a}$, and $Q^{\textsl{c}}_{d,a}$. Every aggregator is assigned with a $\sigma_{d,a}$ value in the range $10-110$.

\begin{table*}[t!]
    \caption{Attributes of the aggregators in the 5-bus system}
    \begin{center}
    \begin{tabular}{ccccccccc}
    \hline\hline
        Bus & Aggregator & $\sigma_{d,a}$ & $\gamma_{d,a}$ & $\mu_{d,a}$ & $P^{\textsl{n}}_{d,a}$ (MW) & $P^{\textsl{c}}_{d,a}$ (MW) & $Q^{\textsl{n}}_{d,a}$ (MW) & $Q^{\textsl{c}}_{d,a}$ (MW) \\ \hline
        2 & 1 & 15 & 11.05 & 0.016 & 84.62 & 42.00 & 25.69 & 13.81\\
        2 & 2 & 85 & 38.68 & 0.045 & 338.49 & 168.00 & 102.78 & 55.22\\
        3 & 1 & 56 & 63.54 & 0.066 & 211.56 & 105.00 & 64.24 & 34.51\\
        3 & 2 & 32 & 45.34 & 0.034 & 211.56 & 105.00 & 64.24 & 34.51\\
        4 & 1 & 100 & 29.99 & 0.089 & 324.39 & 161.00 & 98.48 & 52.92\\
        4 & 2 & 77 & 21.23 & 0.024 & 105.78 & 52.50 & 32.12 & 17.26\\
        4 & 3 & 105 & 10 & 0.087 & 133.99 & 66.50 & 40.68 & 21.86\\ \hline\hline
    \end{tabular}
    \label{aggrdata5}
    \end{center}
\end{table*}

\subsection{Sensitivity Analysis}
The problem is modeled using the Python-based algebraic optimization modeling language, Pyomo \cite{pyomobook} using Python 3.12.3 and solved with the optimizer Ipopt \cite{ipopt} which is based on the interior-point method\footnote{Code available at \url{https://github.com/sachinthyh/social-equity-based-OPF}}. It is initially solved with default SES values for both test systems, and a sensitivity analysis is performed for the 5-bus system by re-scaling all original SESs together from $10-150$\% in $2$\% steps.

\begin{figure*}[t!]
    \centering
    \subfloat[Total satisfaction vs. percentage SES]{%
        \includegraphics[width=0.49\textwidth]{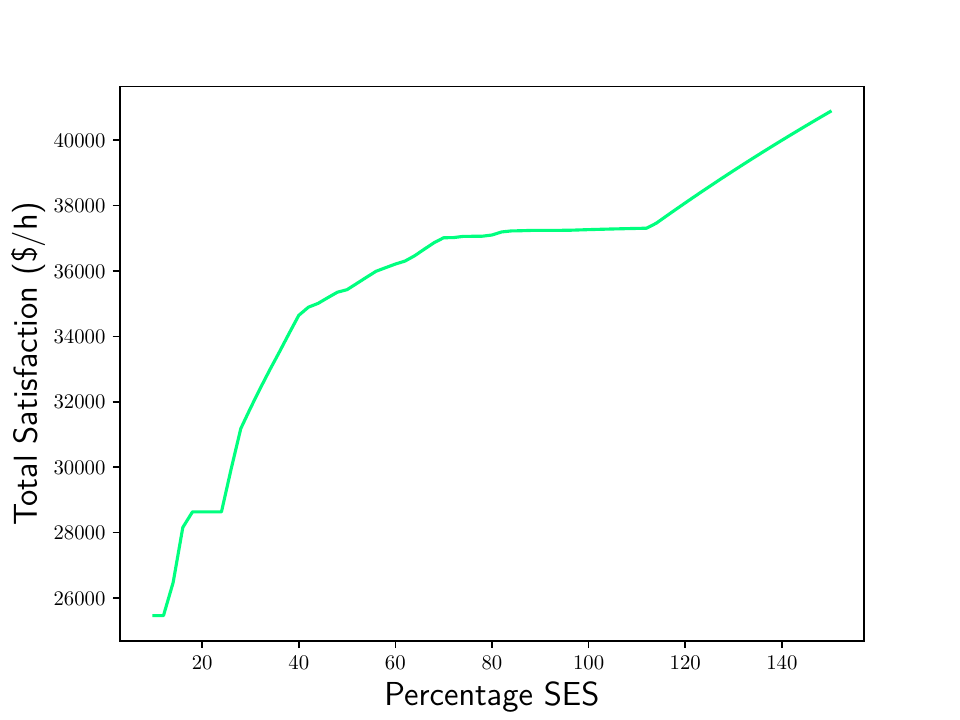}
        \label{tot_util_var}
    }
    \hfill
    \subfloat[Total generation cost vs. percentage SES]{%
        \includegraphics[width=0.49\textwidth]{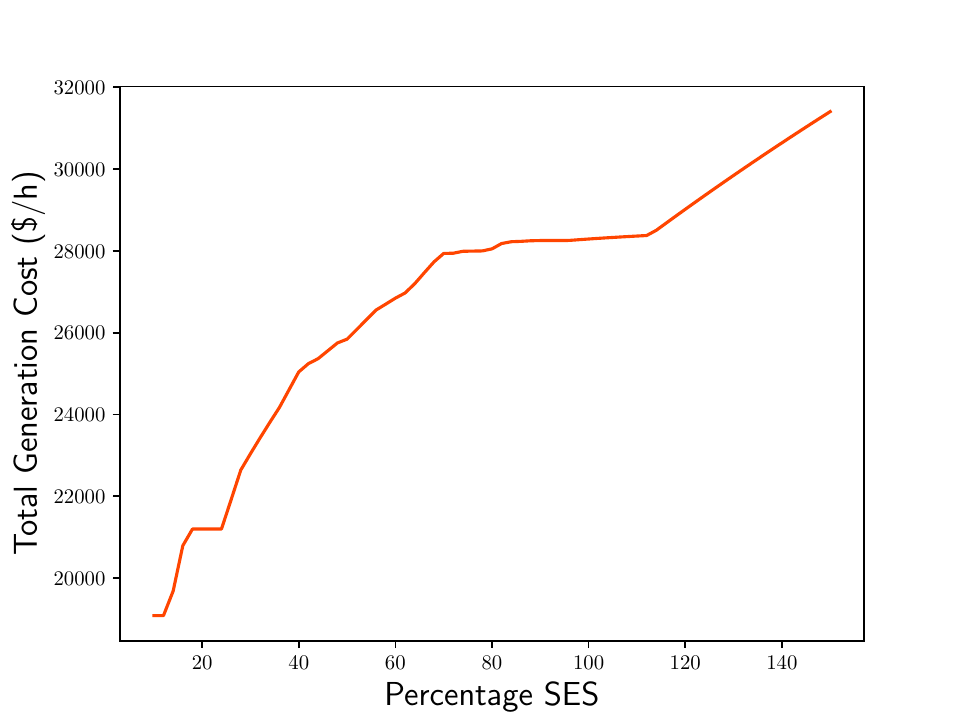}
        \label{tot_cost_var}
    }
    \\
    \subfloat[Social welfare vs. percentage SES]{%
        \includegraphics[width=0.49\textwidth]{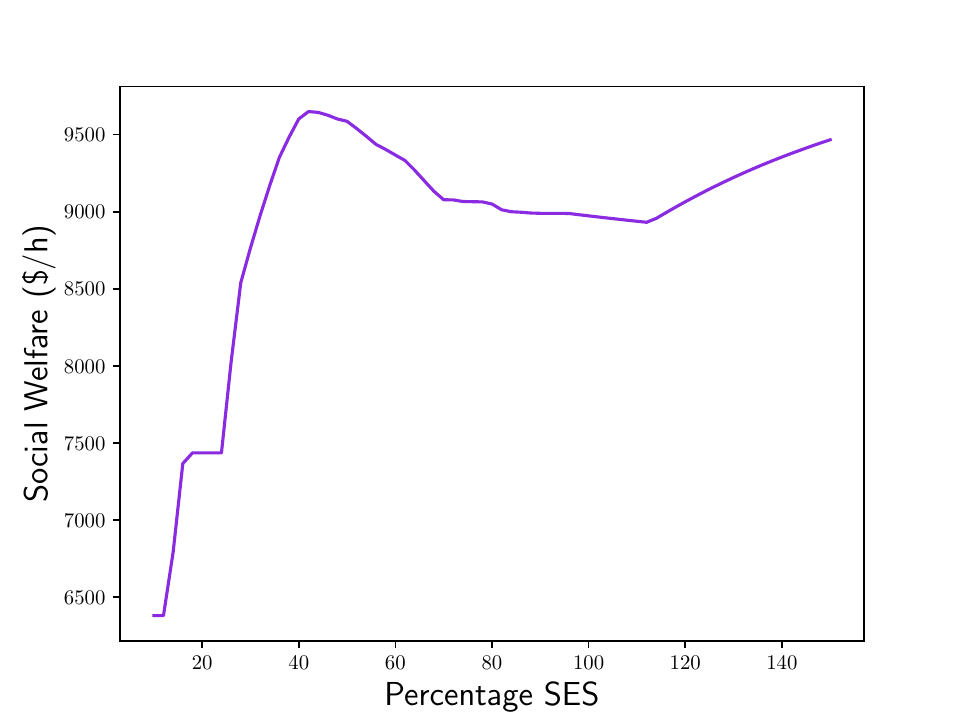}
        \label{sw_var}
    }
    \hfill
    \subfloat[Normalized satisfaction of each aggregator with \%SES]{%
        \includegraphics[width=0.46\textwidth]{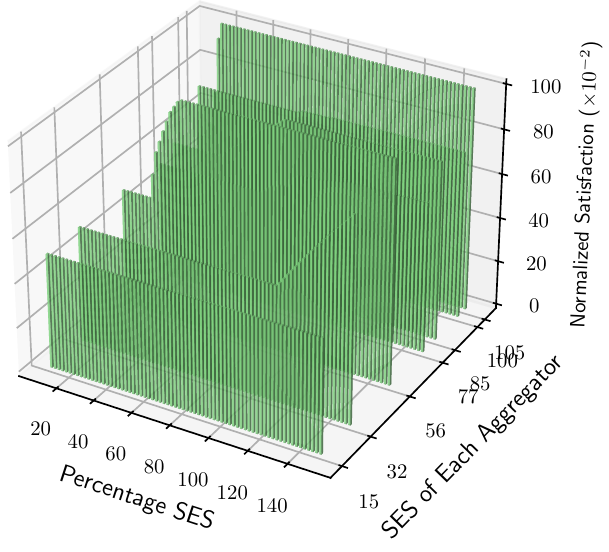}
        \label{indi_var}
    }
    \caption{Results of the sensitivity analysis on the 5-bus system}
    \label{sens_results}
\end{figure*}

\begin{table}[t!]
    \caption{Line ratings of the 5-bus system}
    \begin{center}
    \begin{tabular}{cc}
    \hline\hline
         Line & $S_{l}^{\textsl{max}}$ (MW)  \\ \hline
         $1-2$ & 200 \\
         $1-4$ & 100 \\
         $1-5$ & 120 \\
         $2-3$ & 100 \\
         $3-4$ & 150 \\
         $4-5$ & 120 \\ \hline\hline 
    \end{tabular}
    \label{linedata5}
    \end{center}
\end{table}

\begin{table}[t!]
    \caption{Generator cost data of the 5-bus system}
    \begin{center}
    \begin{tabular}{ccccc}
    \hline\hline
        Bus & Generator & $a_{g}$ & $b_{g}$ & $c_{g}$ \\ \hline
        1 & 1 & 2 & 14 & 60 \\
        1 & 2 & 2 & 15 & 35 \\
        3 & 1 & 2 & 30 & 25 \\
        4 & 1 & 2 & 40 & 20 \\
        5 & 1 & 2 & 10 & 50 \\ \hline\hline
    \end{tabular}
    \label{gendata5}
    \end{center}
\end{table}

\section{Results}
The results of OPF with default SES values are shown in Table~\ref{default_res}. Here, social welfare is the difference between total satisfaction and total generation cost. It reflects the point of optimal trade off between total generation cost and total satisfaction, resulting in a maximized social welfare. The solution is converged with no constraints violated. 
\begin{table}[t!]
    \caption{Results obtained with default socioeconomic scores}
    \begin{center}
    \begin{tabular}{cccc}
    \hline\hline
        System & Satisfaction (\$/h) & Cost (\$/h) & Social Welfare (\$/h) \\ \hline
        5-bus & 37263.06 & 28290.19 & 8972.87 \\
        24-bus & 96848.77 & 77597.78 & 19250.99 \\ \hline\hline
    \end{tabular}
    \label{default_res}
    \end{center}
\end{table}

Fig.~\ref{sens_results} shows the results of the sensitivity analysis on the 5-bus system. Fig.~\ref{tot_util_var} and Fig.~\ref{tot_cost_var} indicates that total satisfaction and total cost are monotonically increasing with the percentage SES. The net total welfare variation, which is depicted by Fig.~\ref{sw_var}, has a peak value around $40$\% of the original SES vales for the scaling interval considered. The normalized satisfaction of aggregator $(d,a)$ can be defined as $\frac{U_{d,a}(P_{d,a})}{U_{d,a}(P_{d,a}^{\textsl{n}})}$. Fig.~\ref{indi_var} shows the variation of normalized satisfaction of each aggregator with percentage SES. One may argue that upscaling the SESs should intuitively lead to increasing total satisfaction, subsequently increasing total generation cost, caused by increased consumption of power by some of the consumers (up to their normal limits, as if the price event was absent), since upscaling implies giving more emphasis to increase satisfaction. This is not always the case, due to the fact that constraints keep this from happening always. This can be explained by regions where total satisfaction and total cost are plateaued. However, when we keep upscaling all SESs together, there may be instances where the consumers of lower relative SESs who were not significantly contributing to the objective, now becomes significant. This is when the plateaued trend becomes an increasing trend again. 
Furthermore, with the goal to maximize social welfare, observation of Fig.~\ref{sw_var} reveals that there could be an further improved solution by re-scaling (up or downscaling) the original SESs. This re-scaling can be incorporated in the model as a multi-stage optimization to improve the solution, iteratively. It is important to note that the shapes of Fig.~\ref{tot_util_var}, Fig.~\ref{tot_cost_var}, and Fig.~\ref{sw_var} are highly scenario-specific and defined by physical system conditions. The computational complexity of this OPF formulation is nearly identical to the conventional OPF formulations, as the objective function is second order, while the constraints are similar to that of the conventional problems.

\section{Conclusion}
Operational planning during price events in modern power systems lack the inclusion of socioeconomic factors to reflect social equity. In this study, we proposed a modified optimal power flow framework to achieve an optimal social welfare while aiming to equitably hedge against the effects of price events. It is intended to incorporate social equity consideration into wholesale electricity markets through a metric called socioeconomic score (SES) that is calculated from the socioeconomic status of the aggregated demand. Results show that, by using suitable weights of SESs, we can maximize the social welfare of the overall system. Future work would focus on refining the weighted consumer satisfaction function to enhance the effectiveness of the model across diverse scenarios.

\bibliographystyle{ieeetr}
\bibliography{NAPS2024_Sachinth_Viththarachchige}

\begin{thebibliography}{10}

\bibitem{eventsintro}
M.~Jain, X.~Sun, S.~Datta, and A.~Somani, ``A machine learning framework to deconstruct the primary drivers for electricity market price events,'' in {\em 2023 IEEE Power \& Energy Society General Meeting (PESGM)}, pp.~1--5, 2023.

\bibitem{vortex}
O.~Baylosis, ``2021 texas extreme cold wave,'' article, ArcGIS, December 2022.

\bibitem{spp}
Committee, ``A comprehensive review of spp’s response to the february 2021 winter storm,'' review, Southwest Power Pool, July 2021.

\bibitem{ferc}
Committee, ``Inquiry into bulk-power system operations during december 2022 winter storm elliott,'' report, FERC and NERC, October 2023.

\bibitem{welfareecon}
Y.-K. Ng, {\em Welfare Economics}.
\newblock Palgrave Macmillan UK, 2004.

\bibitem{swofirst}
M.~Munasinghe, ``A new approach to power system planning,'' {\em IEEE Transactions on Power Apparatus and Systems}, vol.~PAS-99, no.~3, pp.~1198--1209, 1980.

\bibitem{utilfunc}
D.~G. Luenberger, ``Benefit functions and duality,'' {\em Journal of Mathematical Economics}, vol.~21, no.~5, pp.~461--481, 1992.

\bibitem{weber}
J.~Weber, T.~Overbye, and C.~DeMarco, ``Inclusion of price dependent load models in the optimal power flow,'' in {\em Proceedings of the Thirty-First Hawaii International Conference on System Sciences}, vol.~3, pp.~62--70 vol.3, 1998.

\bibitem{swoeqn}
Y.~Ma, W.~Zhang, W.~Liu, and Q.~Yang, ``Fully distributed social welfare optimization with line flow constraint consideration,'' {\em IEEE Transactions on Industrial Informatics}, vol.~11, no.~6, pp.~1532--1541, 2015.

\bibitem{p2p}
H.~T. Doan, T.~H.~B. Huy, D.~Kim, and H.~Kim, ``Fully decentralized peer-to-peer community grid with dynamic and congestion pricing,'' {\em IEEE Internet of Things Journal}, vol.~11, no.~14, pp.~24483--24496, 2024.

\bibitem{econbook}
S.~Stoft, {\em Power System Economics: Designing Markets for Electricity}.
\newblock Wiley, 2002.

\bibitem{nirasha}
N.~Bataduvaarachchi, D.~Alexander, V.~Aravinthan, and A.~Tamimi, ``Emergency load shedding to enhance resiliency considering operator and customer equity,'' in {\em 2024 IEEE Green Technologies Conference (GreenTech)}, pp.~178--182, 2024.

\bibitem{satiscase1}
G.~Li, Z.~Bie, B.~Hua, and X.~Wang, ``Reliability evaluation of distribution systems including micro-grids considering demand response and energy storage,'' in {\em 2012 47th International Universities Power Engineering Conference (UPEC)}, pp.~1--6, 2012.

\bibitem{satiscase2}
E.~Adachi, E.~Brock, and R.~Kravis, ``Understanding price formation in grids transitioning to zero marginal cost generation,'' in {\em 2024 IEEE Power and Energy Conference at Illinois (PECI)}, pp.~1--6, 2024.

\bibitem{sociofac}
J.~Yang {\em et~al.}, ``Developing an area-based socioeconomic measure from american community survey data,'' {\em Fremont, CA: Cancer Prevention Institute of California}, pp.~1--17, 2014.

\bibitem{opfbook}
J.~Zhu, {\em Optimization of Power System Operation}.
\newblock IEEE Press Series on Power and Energy Systems, Wiley, 2015.

\bibitem{adi}
A.~Melagoda, A.-K. Manoharan, M.~A. Rahman, V.~Aravinthan, and A.~Tamimi, ``Optimal restoration of a power distribution system during extreme events considering load criticality,'' in {\em 2023 North American Power Symposium (NAPS)}, pp.~01--06, 2023.

\bibitem{pjm5}
F.~Li and R.~Bo, ``Small test systems for power system economic studies,'' in {\em IEEE PES General Meeting}, pp.~1--4, 2010.

\bibitem{ieee24}
P.~M. Subcommittee, ``Ieee reliability test system,'' {\em IEEE Transactions on Power Apparatus and Systems}, vol.~PAS-98, no.~6, pp.~2047--2054, 1979.

\bibitem{pyomobook}
M.~L. Bynum {\em et~al.}, {\em Pyomo--optimization modeling in python}, vol.~67.
\newblock Springer Science \& Business Media, third~ed., 2021.

\bibitem{ipopt}
A.~W{\"a}chter and L.~T. Biegler, ``On the implementation of an interior-point filter line-search algorithm for large-scale nonlinear programming,'' {\em Math. Program.}, vol.~106, pp.~25--57, Mar. 2006.

\end{thebibliography}

\end{document}